# Mechanical Wannier-Stark Ladder of Diamond Spin-Mechanical Lamb Wave Resonators

Philip Andrango and Hailin Wang

Department of Physics, University of Oregon, Eugene, OR 97403, USA

## Abstract

We report the design and theoretical analysis of Wannier-Stark ladders of diamond Lamb wave resonators that feature mechanical compression modes with ultralow damping rates and host spin qubits with excellent optical and spin properties. The degree of localization in the mechanical Wannier-Stark ladder, which is determined by the ratio of coupling rate to frequency spacing between adjacent resonators, sets the effective range of phonon-mediated coupling between spin qubits. Three nearest-neighbor coupling schemes with distinct geometric configurations and a large range of coupling rates have been developed and analyzed. Additional analysis on the effects of disorder indicates that the proposed Wannier-Stark ladder can be robust against realistic experimental imperfections. The development of quantum networks of spin qubits with long-range connectivity can open the door to the implementation of newly developed quantum low-density parity-check codes in a solid-state system.

I. Introduction

Recent advances in quantum error correction (QEC) codes, especially the development of quantum low-density parity-check (qLDPC) codes, indicate that QEC codes with long-range connectivity can overcome the high overhead of QEC codes that only have the nearest neighbor connectivity, pointing to a promising route toward low-overhead fault-tolerant quantum computers[1-6]. These advances should prompt the development of quantum hardware that features long-range connectivity between qubits. For neutral atom arrays, long-range connectivity between trapped atoms has been demonstrated through dynamic reconfiguration of the atom arrays[7].

Wannier-Stark ladders, a well-known phenomenon in semiconductor physics[8-10], can provide a promising approach to developing on-chip quantum networks of qubits with relatively long-range connectivity. Wannier-Stark ladders and the closely related phenomenon of Bloch oscillations have been experimentally realized in a variety of systems, including semiconductor superlattices[11-13], atoms trapped in an optical lattice[14, 15], photonic waveguide arrays[16], and more recently superconducting circuits[17]. For an electronic Wannier-Stark ladder, an electron is subject to a periodic potential and a constant electric field, leading to the localization of the electron wave function as well as the formation of a ladder of equally spaced energy levels.

A linear chain of mechanical resonators can be employed for the realization of mechanical Wannier-Stark ladders. In analogy to one-band nearest neighbor tight binding models[18-20], a mechanical Wannier-Stark ladder can be characterized by the frequency step or spacing, $F$, and the coupling rate, $\kappa$, between adjacent mechanical resonators. The degree of localization of the mechanical waves, which determines the effective range of the coupling between individual resonators, is determined by the ratio, $\eta=\kappa/F$. The range of connectivity can thus be controlled through suitable choices of the relative values of $\kappa$ and $F$. Relatively long or short connectivity can also be realized in the same mechanical network. In addition to one-dimensional (1D) networks, mechanical Wannier-Stark ladders can in principle be extended to two dimensional (2D) networks.

In this paper, we report the design and theoretical analysis of mechanical Wannier-Stark ladders of diamond Lamb wave resonators (LWRs). A LWR is a thin rectangular elastic plate with free boundaries. Diamond LWRs protected by a phononic band gap shield can feature fundamental compression modes with a GHz frequency and a mechanical linewidth less than 100 Hz at T~7 K [21], and can host spin defects, such as nitrogen vacancy (NV) and silicon vacancy (SiV) centers that have excellent optical and spin properties[22-25]. These spin qubits can effectively couple to strain induced by mechanical vibrations via the orbital degrees of freedom of the color centers[26-28]. As a spin-mechanical system, diamond LWRs can serve as an excellent building block for mechanical quantum networks of spin qubits. Specific schemes for 1D and 2D quantum networks of diamond LWRs have been proposed[29, 30]. More general schemes of mechanical networks of diamond spin qubits for applications in quantum computing and quantum simulations have also been theoretically investigated[31-33]. For a network of spin-mechanical resonators, long-range coupling between two given spin qubits is mediated by mechanical vibrations, or phonons, and can be controlled with processes, such as phonon-assisted (i.e., sideband) optical or spin transitions driven by optical fields [34, 35].

We have developed and analyzed three different schemes to couple two adjacent diamond LWRs. The coupling can take place near the nodes of the two LWRs (NN coupling), between the antinodes of the two LWRs (AA coupling), or between the antinode of one LWR and the node of the other LWR (AN coupling). The three schemes can effectively enable a large range of nearest neighbor coupling rates. Combinations of these schemes can also enable the development of 2D mechanical networks. We have investigated the strain distribution in Wannier-Stark ladders of LWRs and have analyzed the effects of disorders on the behaviors of the mechanical Wannier-Stark ladders. These studies indicate that Wannier-Stark ladders of LWRs with a wide range of connectivity are experimentally feasible.

II. Spin-mechanical Lamb wave resonators

A LWR features both symmetric and antisymmetric compression modes (with respect to the midplane of the plate). For the numerical analysis presented in this paper, we focus on the fundamental compression mode. The maximum displacements, i.e., antinodes of the fundamental compression mode occur at the two short edges of the rectangular LWR, while the node occurs at the line that bisects the LWR and is parallel to the short edges, as illustrated by the displacement pattern shown in Fig. 1a. The frequency of the compression mode is inversely proportional to the LWR length and depends weakly on the LWR width (see Fig. 1b). The frequency is essentially independent of the LWR thickness. For the construction of mechanical Wannier-Stark ladders, we can thus vary the resonance frequency of a LWR by changing its length for a relatively large $F$ or width for a relatively small $F$. Numerical calculations in this paper have been carried out with a COMSOL Multiphysics software package. The diamond parameters used are Young's modulus of 1050 GPa, Poisson ratio of 0.1, and mass density of 3515 kg/m$^3$. A thickness of 0.6 μm is assumed for all the LWRs and connecting bridges.

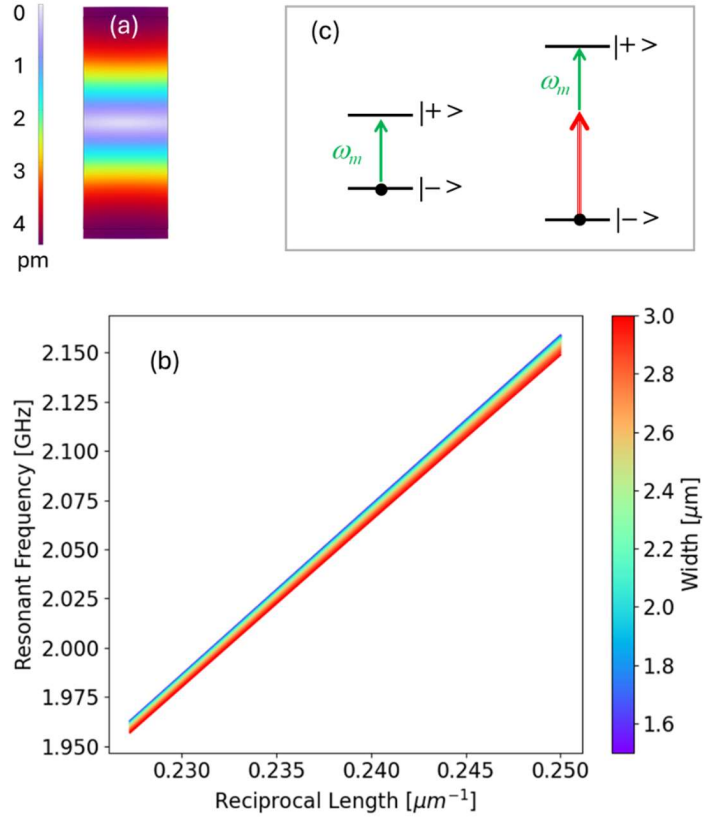

Fig. 1 (color online) (a) Displacement pattern of the fundamental compression mode of a LWR with a relatively large ratio of length over width. (b) The resonant frequency of the fundamental compression mode as a function of the length and width of the LWR. (c) Schematic of a direct acoustic transition between two spin states and a sideband spin transition, which can be driven optically through a Raman transition (not shown).

Spin qubits in diamond can effectively couple to mechanical vibrations through the orbital degrees of freedom. Specifically, strain induced by the mechanical vibrations can result in mixing as well as energy shifts of relevant states. As illustrated in Fig. 1c, strain induced mixing of two spin states can lead to a direct acoustic transition between the two spin states, which has been used for mechanical quantum control of spin states [36]. Strain-induced energy shifts can lead to phonon-assisted transitions, i.e., sideband transitions (see Fig. 1c), including sideband spin transitions driven by optical fields through a resonant Raman process, as shown in earlier experimental studies [34, 35].

Both types of transitions can be employed for phonon-mediated coupling between spin qubits in a mechanical quantum network. For sideband transitions, we can enable and control the coupling between two given spin qubits within the mechanical coupling range by turning on and tailoring the optical driving field for the respective spin qubits. For direct acoustic transitions, spin qubits within the coupling range can all couple to the relevant mechanical modes. In this case, quantum interference techniques can in principle be used for selective coupling between two given spin qubits [37].

III. Coupling between adjacent resonators

To determine the nearest neighbor coupling rate in a mechanical Wannier-Stark ladder, we numerically calculate the rate of coupling (which is half of the normal mode splitting) between two LWRs with the same resonance frequency. The top panel of Fig. 2 shows the displacement patterns of the symmetric and antisymmetric normal modes of the coupled resonators for three different coupling schemes. For the NN coupling scheme, two bridges offset from the node of the compression mode connect the two LWRs. Figure 2a shows the dependence of the normal mode splitting on the bridge length and width. The coupling rate also depends on the offset from the node. We define the offset fraction as the distance between the upper edge of the upper bridge and the lower edge of the lower bridge over the length of the LWR. A smaller offset leads to a correspondingly smaller coupling rate. For relatively short and wide bridges, the coupling rate increases monotonically with increasing width, as expected. However, for relatively long and narrow bridges, the coupling rate can exhibit non-monotonic variations with the width and length. Figure 2a shows, as an example, a strong increase of the coupling rate with decreasing bridge width. This unusual behavior is due to mechanical resonance related to the motion, including the relative motion, of the two connecting bridges.

The AA scheme, for which a bridge connects the short edges, i.e., the antinodes of the two LWRs, features coupling rates that are much greater than those of the NN scheme. In this case, large displacements of the short edges lead to correspondingly large coupling rates between the two LWRs. As shown in Fig. 2b, the coupling rate increases monotonically with increasing width and decreasing length of the bridge. Note that the coupling rate for the AA scheme can reach a large fraction of the LWR resonance frequency. In this case, the overall frequency range of the LWRs in a relatively long Wannier-Stark ladder can approach the resonance frequency of the LWRs involved.

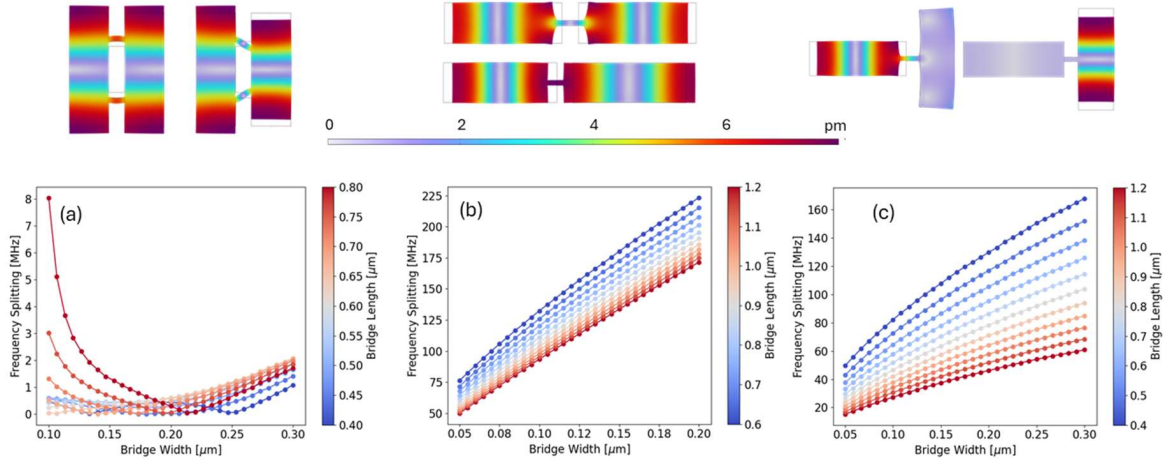

**Fig. 2 (color online)** Normal mode splitting of two coupled LWRs as a function of the bridge length and width. The dimensions of the LWRs are 4.25 μm by 1.5 μm. (a) Node-node coupling scheme, with a bridge offset fraction of 0.15. (b) Antinode-antinode coupling scheme. (c) Antinode-node coupling scheme. The top panel shows the displacement patterns of the corresponding symmetric and antisymmetric normal modes, for which the dimensions of the connecting bridges are 0.8 μm by 0.2 μm.

A more moderate coupling rate between the two LWRs can be achieved with the AN coupling scheme, for which a bridge connects the node of a LWR to the antinode (i.e., short edge) of the other LWR. Again, the coupling rate increases with increasing width and decreasing length of the bridge (see Fig. 2c). Because of its orthogonal configuration, the displacements for the two normal modes occur primarily in one of the resonators, but not both simultaneously, which has important implications for the corresponding 1D network, as will be discussed later. Note that for relatively long connection bridges, the nearest neighbor coupling rate for the AA and AN schemes can also depend sensitively on mechanical resonances of the connection bridge.

The three coupling schemes discussed in this section can also be used for the construction of 2D mechanical networks. Because of its orthogonal geometry of connecting the short edge of a LWR to the long edge of an adjacent LWR, the AN scheme can play a special role in 2D mechanical networks. Combinations of the three schemes with distinct geometric configurations and a large range of coupling rates, spanning from a few kHz to more than 100 MHz, can enable the development of a rich variety of 2D mechanical networks.

IV. Mechanical Wannier-Stark ladders

For a linear chain of mechanical resonators with nearest neighbor coupling rate, $\kappa$, the normal modes of the mechanical system can be described by

$$\kappa(u_{n-1} + u_{n+1}) + \omega_n u_n = \varepsilon u_n \qquad (1)$$

where $\omega_n$ and $u_n$ are the frequency and displacement amplitude of the $n$th resonator ($n = 0, \pm 1, \pm 2, ...$), respectively, and $\varepsilon$ is the normal mode frequency. This equation is the same as that for the 1D tight binding model. For resonators with equal frequency spacing or step, $F$, the solution is the Wannier-Stark ladder, with $\varepsilon_\alpha = \omega_0 + \alpha F$ ($\alpha = 0, \pm 1, \pm 2, ...$). The localized Wannier-Stark states, given in terms of the Bessel function of the first kind, are [18-20]

$$u_n^{(\alpha)} = (-1)^{n-\alpha} J_{n-\alpha}(2\kappa / F). \qquad (2)$$

The degree of localization, which also sets the range of connectivity for the mechanical network, is thus determined by the ratio, $\eta = \kappa/F$.

In this section, we present numerical analysis of mechanical Wannier-Stark ladders of LWRs. Since spin-mechanical coupling takes place through mechanical strain, our analysis will thus focus on the behavior of local mechanical strain, $\delta V/V$, instead of mechanical displacements. Note that maximum strain occurs near the node of the mechanical displacement. We will also analyze the effects of disorders, which are inevitable in experimental implementations, on the Wannier-Stark states.

Figure 3a shows the distribution of mechanical strain in 1D chains of NN-coupled, AA-coupled, and AN-coupled LWRs at a given normal mode resonance. The LWRs all have the same dimensions as those used for Fig. 2 and feature a fundamental compression mode of 2 GHz. For the NN-coupled as well as AA-coupled 1D chains, the sign of $\delta V/V$ alternates across the 1D chain. The AN-coupled 1D chain, however, shows a characteristically different behavior. Relatively strong strain occurs in either horizontally oriented or vertically oriented resonators, but not simultaneously in both types of resonators. For the AN coupling scheme, mechanical displacements in the two directly coupled resonators are orthogonal to each other. In this case, a vertical resonator effectively mediates the coupling between two adjacent horizontal resonators.

Similarly, a horizontal resonator effectively mediates the coupling between two adjacent vertical resonators.

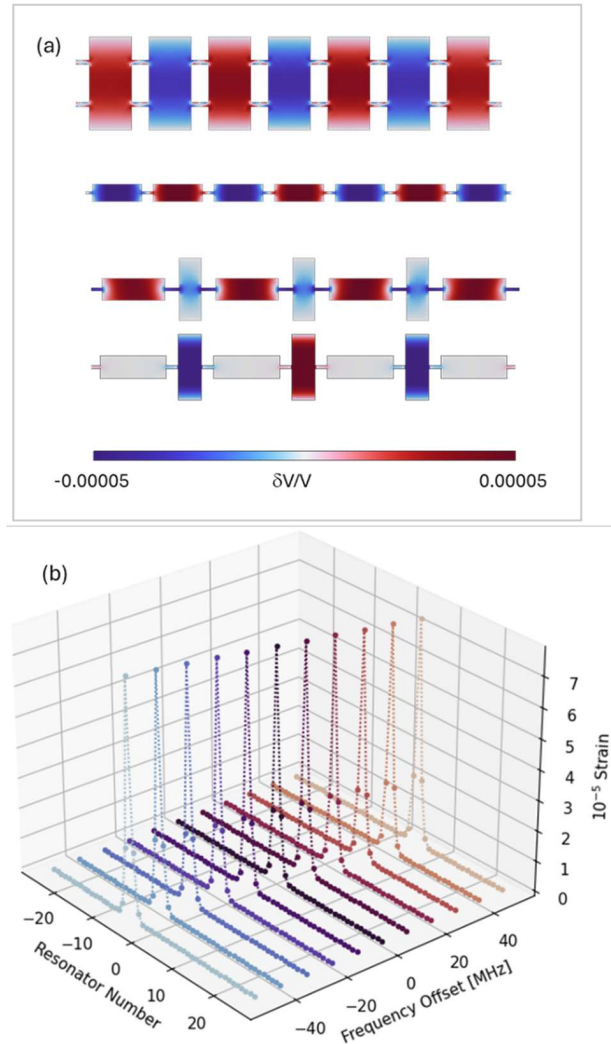

**Fig. 3 (color online)** (a) Distribution of mechanical strain for 1D chains (from top to bottom) of NN-coupled, AA-coupled, and AN-coupled LWRs, all with the same dimensions of 4.25 μm by 1.5 μm, at a given normal mode resonance. The dimensions of the connecting bridges are 0.8 μm by 0.2 μm, with an offset fraction of 0.5 for the NN coupling scheme. (b) Distribution of mechanical strain (absolute value) at the node for a 1D chain of NN-coupled LWRs at a set of normal mode frequencies, with $F=10$ MHz and κ near 2.53 MHz. The dimensions of the connecting bridges are 1 μm by 0.1 μm, with an offset fraction of 0.25.

Figure 3b shows the distribution of mechanical strain at the node for a 1D chain of NN-coupled LWRs at a set of normal mode frequencies, for which the LWR at the center of the chain has the same dimensions as those used for Fig. 2. The frequency spacing or step between adjacent LWRs is set to 10 MHz. As the normal mode frequency shift by one frequency step, the corresponding position of the maximum strain shifts by one LWR, which is a clear manifestation of the Wannier-Stark ladder.

We can achieve the desired range of connectivity for a 1D mechanical network by varying $\eta$. Figure 4 shows the distribution of mechanical strain at the node for 1D chains of NN-coupled LWRs with the same connection bridges and with increasing $F$. The dimensions of the LWR at the center of the 1D chain are the same as those used for Fig. 2. For relatively small $\eta$, the strain decays quickly away from the central LWR, with the connectivity limited to a few resonators. At relatively large $\eta$, the strain can spread over increasing number of LWRs and can exhibit an oscillatory spatial distribution. These states can enable relatively long-range connectivity for the mechanical network. Note that the oscillatory behavior is expected from the Wannier-Stark states given in Eq. 2.

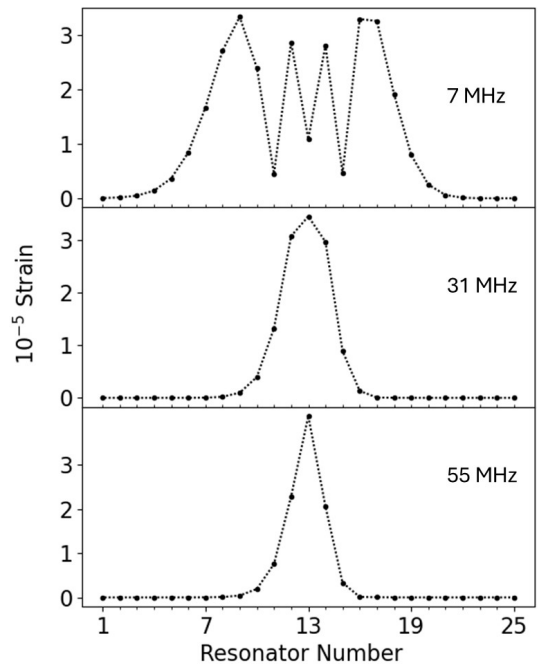

**Fig. 4** Distribution of mechanical strain (absolute value) at the node for 1D chains of NN-coupled LWRs with the same connection bridges and with the frequency step, $F$, indicated in the figure. The dimensions of the connecting bridges are 0.8 μm by 0.24 μm, with an offset fraction of 0.5. The nearest neighbor coupling rate, $\kappa$, is near 21 MHz.

The localization properties of the mechanical normal modes can be further characterized by the inverse participation ratio (IPR), which is defined as [19, 38]

$$IPR(\varepsilon) = \frac{\sum_n |s_n(\varepsilon)|^4}{(\sum_n |s_n(\varepsilon)|^2)^2}. \tag{3}$$

where $s_n(\varepsilon)$ is the maximum strain in the $n$th resonator for a normal mode with frequency $\varepsilon$. For an extended state, the IPR is of order $1/L$, with $L$ being the size (i.e., the total number of LWRs) of the 1D system. The strongest localization corresponds to the maximum possible IPR value of 1. Figure 5a plots IPR for 1D chains of NN-coupled LWRs as a function of the frequency step between adjacent resonators and as a function of the width of the connecting bridges. The dimensions of the LWR at the center of the chain are the same as those used for Fig. 2. The total number of resonators in the chain is 25. The IPR calculations show increasing localization with increasing frequency step between the adjacent resonators and with decreasing bridge width (thus decreasing nearest neighbor coupling rate). The congestion area in Fig. 5a occurs near the onset of oscillatory behavior in the spatial distribution of mechanical strain. The IPR value of the congestion area corresponds to a localization length of 4 resonators, midway between 3 resonators (with no dip in the distribution) and 5 resonators (with one dip in the distribution).

Imperfections in the fabrications of the LWRs and the connection bridges result in fluctuations or errors in both the frequency steps and the nearest neighbor coupling rates. Figure 5b shows the IPR for 1D chains of NN-coupled LWRs that include varying degrees of disorders. For the numerical calculations, we have assumed that the resonator length randomly fluctuates with a Gaussian distribution and with a standard deviation of σ. The IPR value shown is averaged over 50 runs. As can be seen from Fig. 5b, increasing disorder gradually increases the localization of the mechanical normal modes. The overall behavior of the Wannier-Stark states, however, remains largely intact. For the state-of-the-art electron beam lithography, feature sizes as accurate as 10 nm are achievable. In this regard, proposed mechanical Wannier-Stark ladders of diamond LWRs are feasible with the currently available technologies.

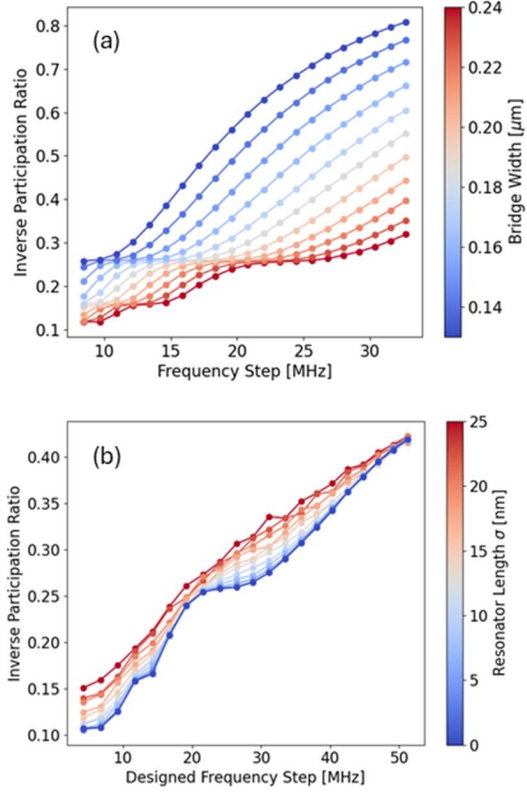

**Fig. 5 (color online)** (a) IPR of 1D chains of NN-coupled LWRs as a function of the frequency step and the bridge width. The bridge length is 0.6 μm, with an offset fraction of 0.5. (b) The same as (a) except that the bridge dimensions are 0.8 μm by 0.24 μm and the resonator length randomly fluctuates with a Gaussian distribution and with a standard deviation of σ.

    A mechanical network can contain multiple Wannier-Stark ladders with a wide range of connectivity. In addition to the NN coupling scheme discussed above, the AN and AA coupling schemes, which feature relatively large κ, can also be used to further increase the range of connectivity. In the limit that the overall frequency range of the LWRs in a mechanical network is small compared with the relevant mechanical resonance frequency, the entire mechanical network can be embedded in a suitably designed phononic crystal and be protected by a phononic band gap. Furthermore, we can in principle reduce or compress the overall frequency range of the LWRs in the network by extending linear Wannier-Stark ladders to zig-zag or sawtooth ladders, which might be necessary when AN or AA coupling schemes are extensively employed in a large mechanical network.

As mentioned earlier, combinations of the three coupling schemes developed in this study can also enable the development of 2D mechanical networks. Strain distributions in 1D mechanical networks shown in Fig. 3a already provide valuable information for employing these schemes in a 2D mechanical network. For example, a vertically orientated LWR can be AN-coupled to four adjacent horizontally orientated LWRs along two orthogonal directions. In this case, the vertically orientated LWR can mediate the coupling between any two of the horizontally orientated LWRs. In this regard, the AN coupling scheme alone can enable a 2D mechanical network.

Based on the properties of 1D Wannier-Stark ladders discussed above, we can anticipate a 2D mechanical network, for which the degree of localization can vary spatially across the network according to a given design, with certain regions featuring relatively short and other regions featuring relatively long range of connectivity. It will be interesting to see if qLDPC codes can be efficiently implemented in this type of 2D mechanical quantum networks. In addition, by extending linear Wannier-Stark ladders to zigzag or sawtooth ladders, we can still protect and isolate a large 2D mechanical network with a phononic crystal band gap shield. Figure 6a shows a 2D network of LWRs embedded in a square lattice of a phononic crystal. The network drawn is only for illustration. The phononic band gap of the square lattice features a large phononic band gap that can protect an entire mechanical network from the surrounding environment, as shown in Fig. 6b.

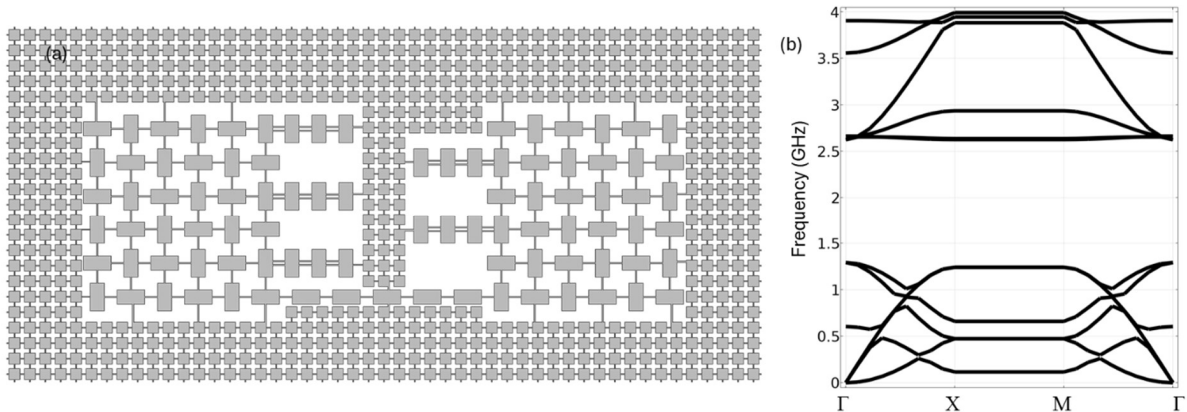

**Fig. 6** (a) A schematic of a 2D mechanical network embedded in a square phononic crystal lattice. The network drawn is only for illustration. (b) Calculated phononic band structure of the square lattice, with a square size of 1.7 μm and with bridge dimensions of 0.7 μm by 0.125 μm, featuring a large and complete band gap centered around 2 GHz.

V. Summary and outlook

In summary, we have designed Wannier-Stark ladders of diamond LWRs, in which spin qubits couple to compression modes via mechanical strain. Three nearest-neighbor coupling schemes with distinct geometric configurations and a large range of coupling rates have been developed and analyzed. The degree of Wannier-Stark localization and thus the range of connectivity can be varied or controlled by choosing suitable coupling rates and frequency spacing between adjacent LWRs. Additional analysis on the effects of disorder also indicates that the overall behaviors of the ladder can remain robust against realistic experimental imperfections. Combinations of the three coupling schemes can also enable the development of a variety of 2D mechanical networks of LWRs.

The mechanical networks of spin qubits discussed this work can in principle feature both long-range connectivity and highly parallel quantum control, which can be enabled by optical spatial multiplexing through optical control of mechanical motion [21] as well as optical control of the spin qubits. Both features are considered to be crucial to the development of large-scale fault-tolerant quantum computers. We hope that our work can stimulate further theoretical efforts in developing qLDPC codes for spin-mechanical systems and can prompt further experimental efforts to exploit mechanical quantum networks of spin qubits for applications in quantum computing. In addition, although our analysis of mechanical Wannier-Stark ladders has focused on diamond-based mechanical resonators, the coupling schemes developed can also be applied to other materials systems, such as SiC, which can also host spin qubits with excellent optical and spin properties[39].

**Acknowledgement**

This work is supported by NSF under Grant Nos. 2012524 and 2003074.